\documentclass[1p]{elsarticle}

\usepackage{graphicx}
\usepackage{color}
\usepackage{soul}

\usepackage{amssymb} 

\journal{Physica A} 

\begin{document}

\begin{frontmatter} 

\title{Memory and long-range correlations in chess games}

\author[famaf]{Ana L. Schaigorodsky}
\ead{alschaigorodsky@gmail.com}

\author[aalto]{Juan I. Perotti}
\ead{perotti@aalto.fi}

\author[famaf,ifeg]{Orlando V. Billoni \corref{cor1}}
\ead{billoni@famaf.unc.edu.ar}

\cortext[cor1]{Corresponding author}

\address[famaf]{Facultad de Matem\'atica, Astronom\'{\i}a y
F\'{\i}sica, Universidad Nacional de C\'ordoba, Ciudad Universitaria,
5000 C\'ordoba, Argentina.}

\address[ifeg]{Instituto de F\'{\i}sica Enrique Gaviola (IFEG-CONICET).}

\address[aalto]{Department of Biomedical Engineering and Computational Science (BECS), Aalto University,
Rakentajanaukio 2, Otaniemi campus, Espoo, FI-00076 Aalto, Finland.}

\begin{abstract}
In this paper we report the existence of long-range memory in the opening moves of a 
chronologically ordered set of chess games using an extensive chess database. 
We used two mapping rules to build discrete time series and  analyzed them using two 
methods for detecting long-range correlations; rescaled range analysis and  detrended 
fluctuation analysis. We found that long-range memory is related to the level of the players. 
When the database is filtered according to player levels we found differences in the 
persistence of the different subsets. For high level players, correlations are stronger at long time scales; 
whereas in intermediate and low level players they reach the maximum value at 
shorter time scales. This can be interpreted as a signature of the different
strategies used by players with different levels of expertise. 
These results are robust against the assignation rules and the method employed in the 
analysis of the time series.  

\end{abstract}


\end{frontmatter} 

\section{Introduction}
\label{}

Some games are complex systems which provide information about
social and biological processes \cite{Prost12,Blasius09PRL,Ribeiro13PO,Sigman10FN,Chassy11PO,
Sire09EPJB,Guerra12PA,Petersen08EPL,Bittner09EPJB,BenNaim07EPL,Heuer09EPJB,Ribeiro12PRE}. 
They are currently being studied because of the existence of very well documented 
registers of games which are useful for statistical analysis. 
In particular, recent works have tried to understand the statistics of wins and losses in 
baseball teams \cite{Sire09EPJB}, the final standing in basketball 
leagues \cite{Guerra12PA}, the distribution of career longevity in baseball \cite{Petersen08EPL},
the football goal distribution \cite{Bittner09EPJB}, and
face to face game rank distributions \cite{BenNaim07EPL}.
The time evolution of the table during a season, for instance,  can be interpreted as a 
random walk \cite{Heuer09EPJB}, and long-range correlations have been found in the score 
evolution of the game of cricket \cite{Ribeiro12PRE}.
Among them, the  game of chess has a main place in occidental culture 
because its intrinsic complexity is viewed as a symbol of intellectual prowess.
Since the skill level of chess players can be correctly identified \cite{Glickman95ACJ}
chess has  contributed to the scientific understanding  of expertise \cite{Chassy11PO}. 
In addition, nowadays, there is a big world-wide community of chess players which 
makes the game a benchmark for studying, for instance, decision making 
processes \cite{Sigman10FN} and population level learning \cite{Ribeiro13PO}.

Exploring a chess database Blasius and T\"{o}njes \cite{Blasius09PRL} observed that the pooled 
distribution of all opening weights follows a Zipf law with universal exponent. 
This is a remarkable result \cite{Maslov09P,nature09} since the Zipf law is followed in 
a range that  comprises  six orders of magnitude. In their work, the authors explain the 
results using a multiplicative process with a branching ratio distribution which resembles 
the real branching distribution measured in the database. Beyond the very good 
agreement between their empirical observations and the proposed model to explain Zipf's law, 
they do not provide an explanation of the evolution of the pool of games; disregarding 
for instance, the question of possible memory effects between different games.
The development of a given chess game should depend on the expertise 
of the players because the first moves (game opening) are usually known in advance 
by high level players as a result of their theoretical background.
Other aspects can influence the development of a chess game according to the level of
expertise.
It has been established that skillful players develop outstanding rapid object 
recognition abilities that differentiate them from the non-expert 
players \cite{Gobet96MC, Gobet96CP}. 
In particular, Gobet and Simon \cite{Gobet96CP} suggested that chess players, like experts 
in other fields, use long-term memory retrieval structures  or templates in addition to 
chunks \cite{Miller56PR,Chase73CP} in short-term memory in order to store information 
rapidly. 
Then, it can be expected that the differences in the level of expertise of chess 
players are reflected in the historical development of chess databases introducing 
correlations between games through memory effects.

Currently there is a big corpus of digitized texts which allows 
the study of cultural trends tracing memory effects \cite{Diuk12FIN}, opening a new branch 
in science known as culturomics \cite{Michel11S}.
In particular, long-range correlations have  been observed in literary corpora \cite{Montemurro02F,Altmann12PNAS}  
where Zipf's law  was  also studied using extensive databases \cite{Petersen12SR}. 
Testing signatures of memory effects is important because it has been shown that systems 
which exhibit Zipf's law need a certain degree of coherence \cite{Cristelli12SR} for its 
emergence. 
Therefore, the detection of long-range memory effects can be useful to shed  
light on the general mechanism behind the Zipf law.

In this work we explored the existence of long-range correlations in game sequences. 
To that end, time-series are constructed using  a chronological ordered chess database 
similar to the one used by Blasius and T\"{o}njes. In order to support the reliability of 
our results, we used two mapping rules to build the discrete time series. 
We also used different techniques to detect long-range correlations,namely the rescaled range
analysis and the detrended fluctuation analysis \cite{Beran94}.

\section{Long-range correlation analysis}

Rescaled range analysis ($R/S$) and detrended fluctuation analysis (DFA) are two useful
tools for detecting long-range correlations in discrete time series. The $R/S$ analysis 
was introduced by Hurst when studying the regularization problem of the Nile 
River \cite{Beran94}, whereas DFA was introduced  by Peng et al. \cite{Peng94PRE} for 
detecting long-range (auto-) correlations in time series with non-stationarities.
Both methods use the accumulated time series which can be thought
as the displacement of a one dimensional random walker whose steps
are dictated by the values of the temporal series. 

As a general procedure, the discrete time series $X(t)$ is first centered and normalized, 
  
\begin{equation}
\label{eq1}
\tilde{X}(t)= \frac{X(t)-\langle X(t) \rangle}{\hat{\sigma}}
\end{equation}

\noindent where $\hat{\sigma}=\sqrt{\langle X(t)^2 \rangle - (\langle X(t) \rangle})^2$ and 
$\langle ... \rangle$ means arithmetic averages over the complete series.
Normalization is useful in order to compare different assignation rules \cite{Montemurro02F}.
Then the accumulated series $Y(t)$ is constructed,

\begin{equation}
\label{eq2}
Y(t)= \sum_{u=1}^{t}\tilde{X}(u).
\end{equation}

Once the series $Y(t)$ is obtained, we applied the rescaled range 
analysis \cite{Montemurro02F,Beran94} and the detrended fluctuation analysis  \cite{Peng92N,Kantelhardt01PA,Bryce12SR}.  
In the $R/S$ method the time series $\tilde{X}(t)$ is divided into non-overlapping 
intervals of size $n$. The range of each interval is computed,

\begin{equation}
\label{eq3}
R(n)= max[Y(1),Y(2), ... ,Y(n)] - min[Y(1),Y(2), ... ,Y(n)],  
\end{equation}

\noindent and the corresponding standard deviation,

\begin{equation}
S(n)=\sqrt{\frac{1}{n}\sum_{i=1}^{n} {(X_{i}-m)}^{2}},
\end{equation}

\noindent where $m$ is the average of $X(t)$ in the interval of size $n$. 
Then the average of the rescaled  range $E[R(n)/S(n)]$ over all the intervals of size $n$ is calculated. 
The Hurst exponent $H$ is obtained varying the size of the
intervals and fitting the data to the expression $Cn^H$ where $H$ is the Hurst exponent.   

In the DFA method, the $Y(t)$ series is also segmented into intervals of
size $n$. At each segment the cumulated series is fitted to a polynomial
$y_n(t)$ and the fluctuation $F(n)$ is obtained: 

\begin{equation}
\label{eq5}
F(n)=\sqrt{\frac{1}{N}\sum_{i=1}^{N} {[y(i)-y_n(i)]}^{2}},
\end{equation}
\noindent where $N$ is the total number of data points. A log-log plot
of F(n) is expected to be linear and if the slope is less than unity it corresponds 
to the Hurst exponent. When $H = 0.5$ the time series resembles a memoryless random 
walker, and for $H > 0.5$ ($H < 0.5$) a random walker with persistent (antipersistent) 
long-range correlations. 

\section{Results}

In our work we use the ChessDB database\footnote{http://chessdb.sourceforge.net/.}
which has $3.5\times 10^6$ registered games; however, all our analysis is based on 
a subset of $1.4\times10^6$  obtained by filtering corrupted data. The total time 
spanned in the database is about $10$ years, from $1998$ to $2007$.
The first step in our analysis consists in mapping the chronologically ordered 
chess games into a discrete time series.
The database contains approximately $380$ games per day uniformly distributed 
over $\approx 3650$ days.
The games played on the same day are not chronologically ordered, 
hence the minimum time resolution in  the studied temporal series is a day.
The mapping for building the time series is far from being unique, but as a 
general rule it should  not introduce spurious 
correlations \cite{Montemurro02F, Altmann12PNAS}. 
We used two assignation rules. 
In the first one, given a subset of consecutive 
games we evaluate the similitude of the last of them with the rest. 
We call this rule the similitude
assignation rule (SAR). We chose to compare the recent $t$--th game 
against the rest using the following mapping expression,   

\begin{equation}
\label{eq6}
X(t)=\sum_{t'=t-\tau}^{t-1}S_d(t,t'),
\end{equation}

\noindent where $S_d(t,t') = k \in \{0,1,2,3,...,d\}$ if and only if the $t$--th and $t'$--th games 
 are identical up to move $k$ included. Here $\tau +1$  is the total number 
of games in the set, and $d$ is the maximum number of moves to be 
considered. In this work we considered full length games, i.e. $d$ is the length 
of the shortest game to be compared. In our database the mean value of a game length 
is $\approx 60$ and the longest game corresponds to $400$ moves. 
According to Eq. (\ref{eq6}) when $\tau=1$ then $X(t)=S_d(t,t-1)$, meaning that each point of 
the time series uses two consecutive  games. For instance, if the sequence of moves of 
two games at $t$ and $t-1$ are --algebraic notation-- 
$e4,e5;Nf3,Nc6;Bb5,a6;...$ and $e4,e5;Nf3,Nc6;d3,d6;...$, respectively, then $X(t)=4$ because
these two games have four common moves.  
In the second rule, $X(t)$ is the popularity at depth $d$ of the t-th game, where the popularity 
at depth $d$ of the t-th game is equal to the number 
of games in the database that have the same moves up to depth $d$. 
We call this rule the popularity assignation rule (PAR) and is similar to the one 
introduced by Montemurro and Pury \cite{Montemurro02F} for the study of a literary corpus. 
The popularity of a given game is computed using its firsts $d$ movements. 
Games that evolve in the same way 
up to a depth $d$ determine a subset of games with the same popularity. 
The popularity of a given game is equal to the number of elements of the subset to which 
it belongs. 
Notice that this assignation rule is based on a global characterization of the database; 
however, it does not introduce spurious correlations as we will see. 

The mapping rules we have chosen are roughly equivalent since games with a popular 
sequence of movess are likely to match with other games. Popular games are related 
to opening lines which develop to different depths. From the chess opening system 
classification it is known that the game lines branch at different depths depending 
on the opening system, for instance, Caro-Kann opening has a long sequence of 
moves before branching. 
However, in a global characerization \cite{Blasius09PRL} the tree of chess moves 
has been proved to be self-similar to a certain degree, which means that the topology
 of the tree is nearly independent of the depth, 
supporting the equivalence of both mapping rules.

\begin{figure}
\includegraphics*[width=12cm,angle=0]{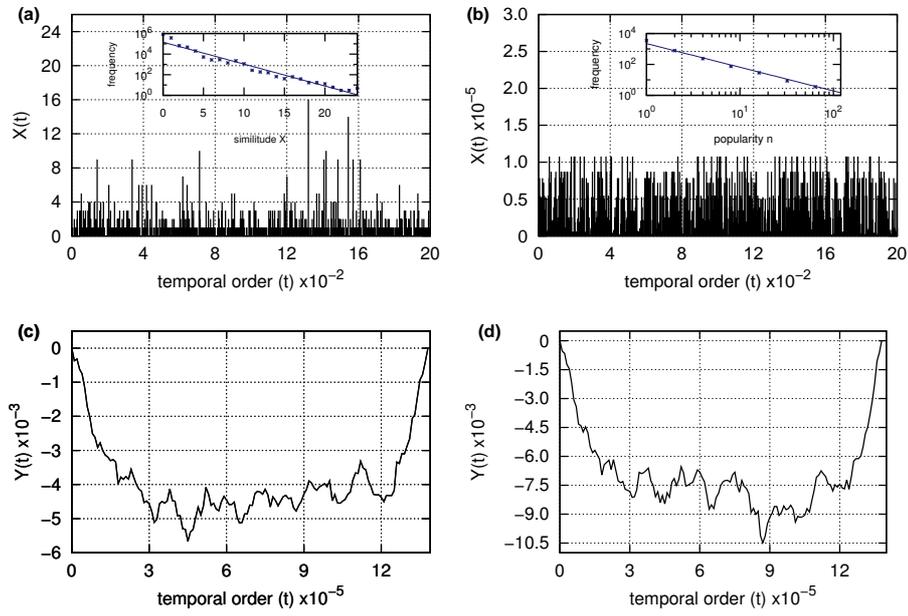}
\caption{\label{fig1} The top panels show fragments of the temporal series $X(t)$
(2000 points) obtained using  the assignation by similitude (a) and popularity (b).
The bottom panels show the complete integrated series, $Y(t)$, 
similitude (c) and popularity (d).  Insets: distribution of similarities (a)
and popularities (b).  The straight lines correspond to fits using: $f(x)= A e^{\alpha x}$
and $g(x)=B x^{\beta}$, insets (a) and (b) respectively, where $\alpha =-0.49$ and $\beta=-1.53$.} 

\end{figure}

The upper panels of Fig. \ref{fig1} show  fragments of the time series obtained using 
the two assignation rules described above. On the right, we show the PAR using a game depth
$d=4$ and on the left  the  SAR using Eq.\ref{eq6} with $\tau=1$.  
These parameters are appropriated for comparing the two assignation rules.
As we will see the mean value of the SAR series (with $\tau=1$) is small, meaning 
that the SAR is testing the first moves, therefore $d=4$ for the PAR is convenient. 
On the other hand, we have checked that the Hurst exponent is nearly independent  of $d$ 
up to $d=20$. For $d>20$ correlations are affected maybe because 
the exponent of the power law which controls the distribution of popularities is much greater
than 2, therefore fluctuations in the popularities become relevant \cite{Blasius09PRL}. 
In the SAR series the Hurst exponent is nearly independent of $\tau$ for  $\tau \leq 500$. 
Beyond that value $H$ drops, and the SAR series do not show long-range correlations. 
Note that in the SAR case, increasing $\tau$ implies a sort of time averaging. 
The points of the series obtained by the PAR are distributed according to a scale free 
distribution as expected. 
In the corresponding inset we show this distribution, the exponent obtained by 
fitting is in agreement with that obtained by Blasius and T\"{o}njes \cite{Blasius09PRL} (see caption). 
SAR series have small values since most of the matches between games are restricted 
to the first moves, in fact the  distribution of coincidences is exponential 
and has a rapid decay (see inset and caption). 
This means that most of the consecutive games are similar in the very first moves, 
and a few of them have coincidences beyond the $10^{th}$ moves.
In the bottom panels we show the complete accumulated series for both cases, SAR on 
the left and PAR on the right. Persistence can be appreciated from the monotonic  
behavior of the curve over long times.

\begin{figure}
\includegraphics*[width=12cm,angle=0]{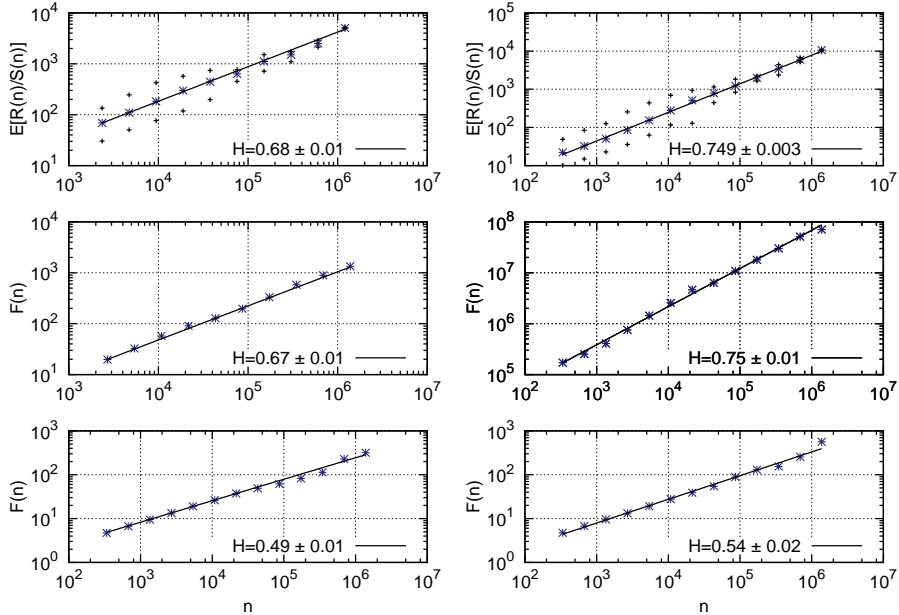}
\caption{\label{fig2} $R/S$ analysis (top panel) and DFA analysis 
(medium and lower panels) for the similitude (left) and popularity  
(right)  assignation rules. The straight dashed lines are linear fits of the data. 
The $R/S$ error points  (black crosses) correspond to the maximum and minimum point of 
the averaged data.}
\end{figure}

In Fig. \ref{fig2}, we show data corresponding to the $R/S$ analysis 
(upper panel) 
and  DFA (medium and lower panels). The left side panels show the analysis of the 
SAR series  and the right one, that corresponding to the PAR series. 
In all cases a log-log plot of the points can be well fitted by a straight line 
as can be appreciated. The Hurst exponent obtained from both analyses $R/S$ and DFA 
are similar; however, the exponent tends to be larger in the PAR series, $H \simeq 0.75$
as compared to the SAR series $\simeq 0.67$. 
According to these values both series exhibit long-range correlations. 
Since SAR is a local assignation rule and PAR is based on a global charactarization of the database, SAR series should be noisier than PAR series. According to this the Hurst exponent of SAR is expected to be less than that of the PAR series.
The fact that both methods give nearly the  same value for the 
Hurst exponent indicates that  non-stationarities are  negligible in the time series. 
When the SAR and PAR time series are randomly shuffled (lower panels), the Hurst exponent 
in both analyses becomes close to $0.5$, which corresponds to series without long-range 
correlations. This means that the assignation rules we used do not introduce 
artificial correlations. 

\begin{figure}
\includegraphics*[width=12cm,angle=0]{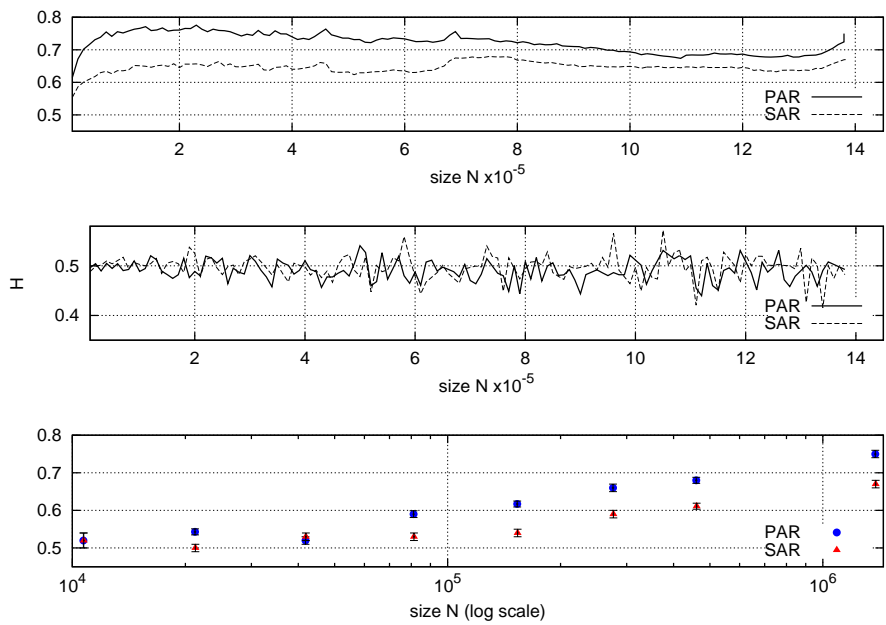}
\caption{\label{fig3} Upper panel: Hurst exponent obtained by the DFA method as 
function of the length of the time series using the two assignation rules, PAR and
SAR series, full and dashed lines, respectively. Meddle panel: Hurst exponent obtained
from the shuffled series. Lower panel: $H$ versus size in the pruned series.
}
\end{figure}

\begin{figure}
\includegraphics*[width=12cm,angle=0]{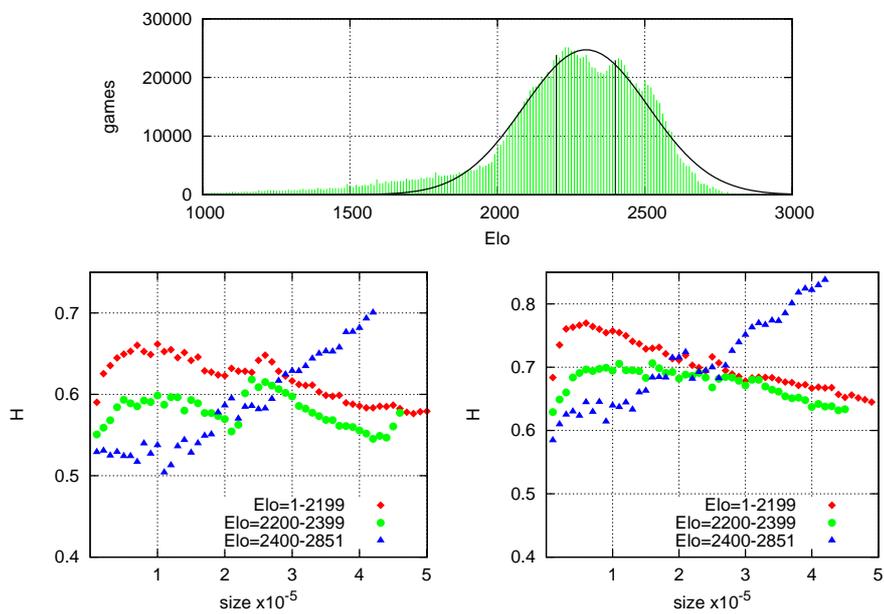}
\caption{\label{fig4} Hurst exponents obtained by the DFA method computed in different
ELO intervals as function of the length of the series. SAR left lower
panel and PAR rule right lower panel. Upper panel:  ELO distribution 
of the whole database, the vertical lines indicates the ELO's intervals.
The continuous line correspond to a fit using a Gaussian function, 
$f(x)=\exp(\frac{-(x-x_0)^2}{2a^2})$, with $x_0=2303$ and $a=203$.}
\end{figure}

In order to compare the correlations obtained from different subsets  of the total 
time series we will first study the dependence of the Hurst exponent on the length
of the series. 
In the upper panel of Fig. \ref{fig3} we plotted the  Hurst exponent as 
function of the length $N$ of the PAR (full line) and SAR (dashed line) series. 
In all this analysis we used the DFA method and we averaged over all 
possible sets of consecutive games of length $N$.
A certain correlation can be appreciated between the Hurst exponent of the two series
assignations. The exponent $H$ of the PAR assignation is always greater than that of the 
SAR assignation, as in Fig.\ref{fig2}. In both cases $H$ grows as the 
length of the series increases reaching a maximum value at nearly $2\times10^5$ 
and then nearly stabilizes to $H\simeq 0.75$ in the PAR series and $H \simeq 0.65$ in the SAR series. 
In the case of the shuffled series (medium panel) we do not observe size effects and 
$H$ fluctuates around $0.5$ as expected. It is worth noting that in this analysis, a change 
in the  size of the time series is equivalent to a change in the total time spanned. 
Finally, in the lower panel of Fig.\ref{fig3} we show the effects of 
reducing the series by pruning them. The error bars are obtained from the fitting errors. 
In this case we constructed reduced series by  removing  points from complete series but  
keeping nearly constant the total time span.
More specifically, to construct the set of reduced series we pick up one point from the 
complete series every $s$ points, where $s = 2, 4, 8, ...$. 
The result after pruning is similar to that observed in the upper panel.
In both cases, the exponent grows as the size of the series increases.
However, in the pruned series $H$ seems not to be stabilized even in the larger series.           
Since the level of the players can be ranked according to their ELOs, 
a rating system introduced by the physicist Arpad Elo \cite{Glickman95ACJ},
we repeated the analysis filtering the database according to different ELO intervals.
We used the following ELO's ranges \cite{Chassy11PO}: $[1,2199]$, $[2200, 2399]$ and above
$2400$. These intervals nearly correspond to: non-experts and master candidates, masters, 
and grand masters, respectively. In particular, this partition
also ensures each interval contains nearly the same number of games.
In Fig. \ref{fig4}, we show the Hurst exponent as function of the length of the time series
associated to the ELO's interval defined above. 
In the upper panel, we show the distribution of ELO's rank of our database.
As expected it is well described by a  Gaussian function $\exp(\frac{-(x-x_0)^2}{2a^2})$ (continuous line),
with the fitted values, $x_0=2303$ and $a=203$.
On the left we show the  SAR series; and on the right, the PAR series. As in Fig. \ref{fig3}, 
the behavior of the two assignation
rules is similar and the exponents of PAR series are  slightly greater than in the SAR series. 
At short times the Hurst exponent in the three ELO intervals is close to 0.5,  
i.e. at this time scale games are uncorrelated and no persistence can be detected. 
At intermediate times, the lower ELO intervals show stronger correlations; 
and at the longest times, the order reverses and the games of high level players are 
the most correlated.

\begin{figure}
\includegraphics*[width=12cm,angle=0]{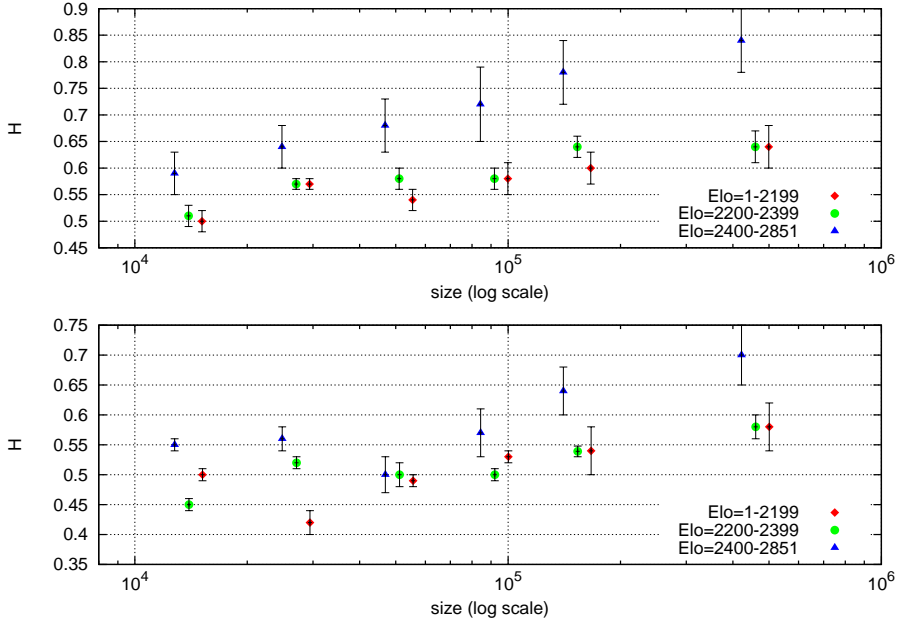}
\caption{\label{fig5} Size effects in the Hurst exponent for pruned series in 
different ELO ranges. Upper panel PAR series and lower panel SAR series.
}
\end{figure}

When a pruning is implemented in series filtered according to ELO ranks,
the Hurst exponent decreases as the length of the series decreases, as in the 
complete database (see Fig.\ref{fig5}). The error bars are obtained from the fitting errors.
At variance with the results of Fig.\ref{fig4} 
the series of higher ELOs are always more correlated, suggesting a different 
origin of this behavior. 

\section{Summary and Discussion}
\label{conclu}

The Hurst exponents obtained through $R/S$ and DFA analysis are similar 
indicating that the time series analyzed are stationary for both assignation rules,
SAR and PAR. 
For the case of PAR and the $R/S$ analysis our results are similar to that reported
in literary corpora \cite{Montemurro02F}.
The value $H\simeq0.5$ that emerges from the shuffled series implies that 
the assignation rules we used to build the series do not introduce spurious correlations.
The analysis of the series as a function of the time spanned shows that there 
is a threshold in the length of the time series from which long-range correlations 
appear. This threshold depends on the assignation rule and on the level of 
the players.  
When the database is filtered by ELO ranks, our results indicate that long-range 
correlations observed at large time scales are related to the presence of high level 
players. The games corresponding to intermediate and low level players show correlations 
at shorter times  when compared to high level players. 
A reduction of the database by pruning indicates that in short series correlations disappear. 
At variance with the analysis based on the length of the time spanned, the reduction in 
correlations by pruning has a similar behavior irrespective of the ELO ranks analyzed. 
Therefore, the reduction  of correlations in this case should have a different origin. 

Size effects in the detection of long-range correlations have been analyzed in the work 
of Coronado {\it et al.} \cite{Coronado05JBP} using time series generated by the Makse 
algorithm \cite{Makse96PRE}. They observed that when DFA is used, size effects are 
negligible even in series containing as few as $10^3$ points.
Since we use DFA and in our series size effects were observed 
in series having more than $10^3$ elements, 
our results must be reflecting intrinsic properties of the system.
However, in our case we have portions of the time series which are randomly shuffled due 
to the minimum time resolution of our database, and this is affecting the detection 
of persistence in the time series.
In long-memory analysis of persistence if missing data are  replaced using random data drawn from 
the empirical distribution of the time series, the Hurst exponent tends to decrease \cite{Wilson03PRE}.    
Notice that pruning  does not affect the total time  spanned in the series.
When pruning, the parameter $s$ can go up to $s \approx 380$ and still have on average 
one game per day in the pruned series with a time series of length $\approx 3,650$. 
According to this, the pruned time series we used should be representative samples of 
the temporal evolution of the games.
However, as was pointed out, in pruned series correlations decrease.
In our case, a possible explanation is that pruning affects the representativity of samples because 
persistence is a collective effect. 
The samples in order to be representative have to include at each time the behavior of a 
representative pool of different players.
When series are pruned, the number of samples taken at the minimum resolution time (one day) are 
reduced hence the behavior of the pool of players is not correctly probed. 

For intermediate and low level players persistence reaches its maximum in two or three years 
and then nearly stabilizes. This behavior is independent of the origin of time used 
for the analysis. In fact, the points shown in Fig. \ref{fig4} result from the averaging 
of equivalent non-overlapping time intervals. For the most specialized players a signature 
of persistence appears after one or two years. According to our results, it seems that 
very high level players use different strategies as compared to low level players. 
It seems they vary the lines of play more often at short times scales, maybe because outstanding players know many opening lines in depth and are less influenced
by the opponents. 
In particular, in high level players the correlations at large time scale are the strongest.

This work was partially supported by grants from CONICET, SeCyT--Universidad 
Nacional de C\'ordoba (Argentina), and the Academy of Finland, project no 260427. 
The authors thank Pedro Pury for useful discussion. 

\bibliographystyle{elsart-num}

\end{document}